# ABOUT THE SUITABILITY OF CLOUDS IN HIGH-PERFORMANCE COMPUTING

Harald Richter

Clausthal University of Technology, Clausthal, Germany
hri@tu-clausthal.de

*ABSTRACT*

*Cloud computing has become the ubiquitous computing and storage paradigm. It is also attractive for scientists, because they do not have to care any more for their own IT infrastructure, but can outsource it to a Cloud Service Provider of their choice. However, for the case of High-Performance Computing (HPC) in a cloud, as it is needed in simulations or for Big Data analysis, things are getting more intricate, because HPC codes must stay highly efficient, even when executed by many virtual cores (vCPUs). Older clouds or new standard clouds can fulfil this only under special precautions, which are given in this article. The results can be extrapolated to other cloud OSes than OpenStack and to other codes than OpenFOAM, which were used as examples.*

## 1. INTRODUCTION

Cloud computing has become the new ubiquitous computing and storage paradigm. Reasons are e.g. the pay-as-you-go accounting, the inherent elasticity, flexibility and user customizing. An example therefore is the arbitrary creation and deletion of virtual machines (VMs) with individual numbers of vCPUs and flexible amounts of virtual main memory per VM. Companies, institutions and individuals can profit from this paradigm by off-loading computing and storage to commercial cloud service providers (CSPs), because CSP services range from simple data backups to entire virtual data centres. These advantages make cloud computing also attractive for scientists, since they do not need any more to provide and maintain their own computer infrastructure, but can out-source it to CSPs, which is then hosted as virtual IT. However, for the case of High-Performance Computing (HPC) in a cloud, as it is needed in simulations or Big Data analysis, things are getting more intricate, because HPC codes must stay highly efficient and thus scalable, even when executed by many virtual cores (vCPUs), which are located on different physical servers. This is not necessarily the case in older clouds and also not in newer standard clouds, as measurements made by several research groups have shown. For example, the US Dep. of Energy (DoE), which is responsible for HPC in the USA, has questioned the usefulness of clouds for HPC in 2011 [1] in general. Other authors with the same opinion are [2]-[4], for example. We believe that further research effort is needed to improve the execution efficiency and the speed-up for HPC in clouds, but it is also our opinion that this is possible. In this contribution, multiple reasons for cloud inefficiencies are presented for the case of OpenStack [5] as cloud operating system and for OpenFOAM [6] as HPC example code. Suggestions are made how to solve or circumvent them. The results can be extrapolated to other cloud OSes and other HPC codes. The contribution is organized as follows: In chapter 2, the state-of-the-art is reviewed. Chapter 3 describes our project and what equipment and tools we were using. In chapter 4, the conducted measurements and findings are presented and discussed. The paper ends with a conclusion, followed by an outlook and a reference list.

## 2. STATE-OF-THE-ART

Several HPC-related scientific projects dedicated to cloud optimization and scheduling were studied by us, and a thorough literature search was conducted to determine which characteristics of the cloud are needed to make it HPC-capable. The papers, that are reviewed here are [3], [4] and [7]-[11] (in rank of importance as we felt it). Also the author of this paper has worked previously on cloud performance issues in [12] and in [14]. In [3], [4] and in this article, it is stated that the overhead in virtualized communication can lead to bad performance because of inter-server bottlenecks. This is true if not the latest hardware-accelerators for virtualized communication are used. Furthermore, according to the studied literature, existing cloud schedulers ignore the needs HPC codes have, as well as the heterogeneity and the multi-tenancy clouds have with respect to their resources. It is furthermore said that these are the major cloud-intrinsic bottlenecks that prevent from effective HPC.

**Scheduling:** In order to address the HPC bottlenecks, the authors of [3], [4], for example, have enhanced the Nova scheduler of OpenStack with the information that a job is a HPC application. Together with other meta-information for Nova, they achieved a performance improvement of 45%, which is a remarkable result. In [7], a monograph on scheduling approaches is given, which is also relevant for clouds. In [8], inter-cloud meta-schedulers are discussed, that consider cloud system dynamics, interoperability and heterogeneity issues. The intention of the authors is to elicit the characteristics of a given HPC code and to produce from that information a model that reflects the resource requirements of the HPC job in so-called cooperative e-science infrastructures. In [9], it is reported that schedulers in Hypervisors, such as in XEN, cannot handle adequately heterogeneous workloads from high and low performance compute jobs at the same time. The reasons for that is that inter-VM communication inside of the same HPC job is degraded by the fact that a VM can be descheduled in the very moment of their communication with another VM that is not descheduled. As a result, both cannot proceed. The authors suggest to schedule HPC jobs not on a cloud-wide basis, but only inside of isolated subsets of cloud resources to limit inter-job interferences. The authors are using a predictor model and a software implementation of it for a prognosis, which VM will communicate with which other, in order to avoid descheduling at the wrong point in time. Furthermore, they migrate a communication-intensive group of VMs to another resource subset, in order to make IO-dominant HPC more effective. This is accomplished by a scheduler that is aware of the IO activities of VMs, i.e. of their ongoing communication relationships.

**VM Placement:** In article [4], the idea of a better placement of VMs by Nova is deepened. The authors have modified Nova, in order to make it aware of the underlying cloud hardware, of the topology of the interconnect network, of the arrangement of resources and of interferences between jobs because of noisy neighbours.

**Throughput vs User Satisfaction:** In [10], it is reported about a distributed job management system that is supposed to support millions of small HPC jobs. This system aims to the big commercial CSPs such as Amazon and Google. The focus lies on high throughput and good utilization, which is exactly what CSPs need, but not in minimizing the elapsed time for individual HPC jobs. In [11], the term HPC-as-a-Service is introduced as a new offer from CSPs. The project tries to bridge the gap between what a CSP can offer as compute resources to his clients at a specific moment in time and for a specific price, and what clients want to have at that moment. It is explained that both sides (clients and CSPs) exhibit big variances and heterogeneities. It is furthermore pinpointed that a multi-criteria optimization of cloud resources is needed because of that. The authors used for that purpose mixed integer linear programming and a stochastic optimization model for efficient HPC resource sharing for service provisioning. Their focus lies on the cost-benefit of cloud resources.

**Trustfulness of Results:** The papers we have reviewed have contributed to HPC-efficient clouds. However, some papers were using not a real cloud, but some simulator, or they have not used a real HPC code, but synthetic load generators. From our point of view, it was not always clear how realistic the achieved results are. Therefore, we followed the path of real hardware executing a widely-used HPC package, and to make with this package real measurements on a standard cloud in the hope to achieve more realistic results.

## 3. PROJECT DESCRIPTION

For our project, we built an own cloud, installed OpenStack and used OpenFOAM as HPC benchmark, which is based on the MPI parallelization standard [13]. Furthermore, shell scripts were written to automate the OpenFOAM benchmarks by running them with various parameters and set-ups.

### 3.1. Easier Possibilities

Before we started to establish an own cloud, we have investigated the subsequent easier possibilities:

1.) Installing a cloud on a set of VMs (nested virtualization)
2.) Using a cloud simulator
3.) Doing all measurements in a commercial cloud or on a University cloud in a computing centre.

All three options were evaluated, and it is explained in the following why we dismissed them all together.

**Nested Virtualization:** We found out that already a single virtualization that is not nested, decreases HPC speedup and efficiency, unless the latest available hardware accelerators for virtualized computation and communication are engaged, which are described in [14], for example. To install OpenStack on a set of VMs in order to create VMs insides of VMs was therefore not an option. Cloud Simulators: From the set of easily available open-source cloud-simulators, we started with a closer look to CloudSim [15]. The alternatives we have considered before and dismissed were GreenCloud [16], iCanCloud [17] and an improved version to the MaGateSim simulator, which is described in [18]. GreenCloud and MaGateSim are for energy-saving cloud-computing, which is not in our focus. From our point-of-view, they were too limited for the required performance analyses. Furthermore, iCanCloud helps to predict the trade-off between cost and performance, which is also not relevant for us. A closer look into CloudSim revealed that it contains a very limited model for communication, which cannot reflect sufficiently the MPI-based communication of OpenFOAM, and also not the complex virtualized communication of OpenStack. For example, it does not model separately inter-core, inter-processor and inter-server communication. CloudSim alone is too restricted for what we need. Additionally, we found the paper [19], which stated that the results from CloudSim are not realistic. Because of that, the authors of [19] created a substantial update called NetworkCloudSim, which supports a more advanced bandwidth/latency model. We concluded that from all network simulators only NetworkCloudSim has relevance for us. It is in principle possible to profit from NetworkCloudSim by a simulative exploration of models for cloud applications. These models are typically defined in terms of estimated job duration and communication times between the parallel tasks of a job. However, the claim of the authors that NetworkCloudSim allows for precise evaluation of scheduling algorithms in scientific, MPI-based applications, including the modelling of a data centre's interconnection network could not be verified by us. The problem with this claim was that the authors did not provide any figures or examples from real measurements to gauge the many NetworkCloudSim parameters. Furthermore, hardware accelerators, such as Single Root IO-Virtualization (SR-IOV) [14], [20], for example, which are nowadays indispensable for an efficient virtualized computation and communication, are not contained in NetworkCloudSim. They must be implemented by the user with high effort. Furthermore, no predefined model exists for inter-VM communication via the KVM Hypervisor or for the OpenVSwitch [21] of OpenStack. Such a model would have been very hard to realize by the user, because the virtual networks in OpenStack follow the principle of Software-Defined Networking (SDN), with the result that they are dynamically variable over time, which is a challenge for every model. Furthermore, NetworkCloudSim does not provide a ready option to model the influence of tunneling protocols, such as GRE [22], and of VLAN or VXLAN [23]. We cannot ignore them, since they are frequently used in clouds. Finally, due to our literature search, nobody else has modelled so far in NetworkCloudSim the distinct configuration parameters of a Hypervisor, of a hardware accelerator, of VLANs or of tunneling protocols. This meant for us that it was not possible to obtain realistic results without tremendous own efforts for software development and parameter gauging: NetworkCloudSim does not provide ready options to model the communication structure and setup of an HPC application reliably enough. This made a real cloud indispensable for us. The only question was, whether an own cloud or an alien cloud would be the easier solution. Existing Commercial or University Clouds: We learned quickly that it is not possible to change on-the-fly the interconnect structures in a commercial or University cloud, or to add contemporary hardware accelerators, because this disturbs productive operation. Furthermore, system administrator rights would have been needed, which cannot be obtained for an alien infrastructure. Further problems have been that no CSP known to us allows for specific placements of user VMs in his computing centre in order to influence deliberately the other loads, which co-exist at the same time. Because of that, it was not possible for us to exclude measurement errors caused by the Noisy Neighbour problem. Neither could we ensure this way the reproducibility of the measurement results. For that reasons, we decided that all three options discussed above are not viable, and we decided to establish an own cloud.

### 3.2. Our Project Cloud

Our cloud consists of 17 used servers from Dell and Sun, which were at the time of the measurements older than 4 years. We had a total number of 76 Cores, 292 GB RAM and 19 TB as Disk Storage. The servers were coupled by 17 Infiniband network interface cards of 40 Gbit/s each and a 40 Gbit/s Infiniband switch. The interfaces are of type Mellanox MHQH19B-XTR and are using QSFP copper cables, as well as the switch itself, which is of type Mellanox Infiniscale IS5023. The switch has 18 ports and a low port-to-port latency of 100 ns only. In parallel to that high-speed network, a standard communication system was installed, that comprises 17 Ethernet cards of 1 Gbits/s each and a Ethernet switch with 24 port of 1 Gbits/s, respectively, to allow for performance comparisons between the two couplings. The host OS for the cloud was Ubuntu 14.04.01 with the OpenStack IceHouse release installed, while Ubuntu 12.04.05 was used throughout as Guest OS.

### 3.3. Integration of Infiniband in Our Cloud

For the integration of Infiniband in our cloud, the virtual Ethernet-network interface-cards (NICs), which are the standard API of KVM for the user, were realized by us by means of the TAP device driver [24]. TAP simulates a NIC by software, and users communicate via TAP by read/write file operations. These operations are translated by TAP into payloads for virtual Ethernet frames, which are subsequently forwarded by OpenVSwitch by means of L2 switching. OpenVSwitch is as KVM an important component of OpenStack. Both are initialized and configured by OpenStack. After that step, OpenVSwitch provides for every TAP a virtual switch port, at which TAP can feed-in its virtual Ethernet frames. Additionally, OpenStack and KVM provide for the VMs of every customer an own VLAN, which is isolated from the VLANs of other customers, in order to provide for IT security. OpenVSwitch processes each virtual Ethernet frame such that the frame is either delivered by means of the user VLAN to a VM in the same server, or alternatively, such that a switch output-port forwards the frame via a GRE tunneling protocol to another Host OS. Since the transportation of an IP packet in an Infiniband Frame is not possible, the Infiniband-over-IP (IPoIB) protocol [25] was added as carrier. We have found no other possibility to integrate Infiniband in OpenStack without SR-IOV.

### 3.4. Our HPC Application

As an example for a HPC application, the widely used OpenFOAM was selected, which is based on Open MPI [26]. It is a parallel HPC code for the numeric solution of Laplace and Navier-Stokes equations, i.e. for the calculation of laminar and turbulent flows of compressible and incompressible fluids, which are gases or liquids. OpenFOAM has additional solvers for general particle flows, for combustion, molecular dynamics, heat transfer, electromagnetic problems, solid elastic bodies, and other purposes, which were not used by us. The reason for the latter was: before we started with OpenFOAM, we performed a questionnaire by asking users of OpenFOAM what they are exactly doing, and what their expectations are when executing the code on a cloud. From that questionnaire, we understood that OpenFOAM is mainly used to solve the Navier-Stokes equations, and we learned also that users considered OpenFOAM and OpenStack as unfavourable combination, unless the latest server generation is used in the cloud.

## 4. PERFORMANCE TESTS

Initially, we configured OpenFOAM to execute the Dam Brake example that comes with the 2.2.1 distribution, because it is well documented. In this example, there are 7700 grid points for geometric objects in two dimensions. One second in reality is simulated by 1000 time steps. After a first simulation run, we modified the initial configuration to make more advanced tests. All measurements were repeated 50 times, and the first run in each measurement cycle was deleted in order to exclude transient effects.

### 4.1. Measurement Results

The execution-time results of all test set-ups are shown in table 1. The results were post-processed by calculating the speed-up and the efficiency of the cloud. These two metrics are defined by:

Definition 1. Speed-up S is the ratio of the execution times of a sequential code before and after virtualization and parallelization.

Definition 2. Efficiency E is the utilization of n server cores or OpenStack vCPUs and defined as E = S/n.

Measurement results for the set-ups 1-7.

| Set-Up | Wall-Time [s] | Speed Up | Efficiency [%] |
| --- | --- | --- | --- |
| 1a: 1 core, bare metal | 144 | 1 | 100 |
| 1b: 4 cores, 2 CPUs, 1 server, bare metal | 46 | 3.1 | 78 |
| 2a: 1 core, 1 KVM | 180 | 0.8 | 80 |
| 2b: 4 cores, 2 CPUs, 1 server, 1 KVM | 62 | 2.3 | 58 |
| 3: nested virtualization, 1 core, 2 KVMs | - | - | - |
| 4a: 1 core, 1 KVM, OpenStack | 154 | 0.94 | 94 |
| 4b: 4 cores, 2 CPUs, 1 server, 4 KVMs, OpenStack | 60 | 2.4 | 60 |
| 5: 4 cores, 4 CPUs, 4 servers, 4 Ethernets, 4 KVMs, OpenStack | 320 | 0.45 | 11 |
| 6: 16 cores, 4 CPUs, 4 servers, 4 Ethernets, 16 KVMs, OpenStack | 670 | 0.21 | 5 |
| 7a: 4 cores, 4 CPUs, 4 SUN servers, 4 Infinibands, 4 KVMs, OpenStack | 237 | 0.61 | 15 |
| 7b: 16 cores, 4 CPUs, 4 SUN servers, 4 Infinibands, 16 KVMs, OpenStack | 998 | 0.14 | 4 |

## 4.2. Evaluation of Measurement Results

According to set-up 1a, the reference execution time for all subsequent measurements was determined as 144 s. This value indicates that the problem size compared to usual HPC execution times is too small, although it is the standard example of OpenFOAM. From set-up 1b, it can be seen that parallelization is beneficial in our cloud if the code execution takes place in the same server. However, efficiency already drops by 22 percentage points to 78% on 4 cores. On a supercomputer or a parallel computer, OpenFOAM should scale well until about 1000 cores, according to its manual. A drop of 22 points already at 4 cores is a sign that the communication time cannot be neglected compared to the computation time. It confirms that the used problem size is too small. From set-up 2a, it can be seen that virtualization causes the efficiency to drop by 20 percentage points. This can be explained by the fact that only the relatively old AMD V [14] accelerator in the CPUs was used, but not newer methods, which could reduce better the efficiency losses that are caused by virtualization. Set-up 2b shows that the simultaneous usage of virtualization and parallelization reduces efficiency by 42 percentage points, which could be expected already by adding the figures 1b and 2a. Set-up 3 was not possible to conduct, because the VM that was created by KVM inside of another VM -according to nested virtualization was not able to run its guest OS. The reason for this is unknown. Set-up 4a shows that the cloud OS incurs an overhead, such that the efficiency drops by 6 percentage points which is low. However, Set-up 4b shows an efficiency drop of 40 percentage points for the parallel code. In Set-up 5, a drastic drop down to 11% can be observed in case of code distribution over 4 vCPUs, which are residing on 4 different servers, which is not tolerable for HPC. In set-up 6, the situation escalates to 5% efficiency, when the code is executed in parallel on 16 vCPUs from 4 servers. Finally, the biggest surprise to us was the measurement in setup 7a, because efficiency increased only to 15%, although Infiniband with the 40-fold data rate was used instead of 1 Gbits/s Ethernet. In setup 7b, Infiniband is even worse than Ethernet in setup 6, which is remarkable. In the following, it is explained how this has happened.

### 4.3. Communication Overheads

We explain the surprising behaviour of Infiniband by the following facts: 1.) the payload of the Infiniband network in setup 7b is shorter than in setup 7a, because the same problem size is divided by 16 vCPUs instead of 4. Shorter payloads, however, increase the Host OS overhead and thus decrease efficiency, because the Infiniband header remains the same. 2.) the minimum transport unit Infiniband can carry is 256 Bytes, while Ethernet needs only 64 Bytes. However, as soon as the problem size gets too small, not enough intermediate computational results can be exchanged between grid borders. As a consequence, more padding bytes are needed in case of Infiniband than for Ethernet. 3.) Infiniband was integrated by us by means of several additional device drivers and protocols, because without SR-IOV there was no other possibility. As a consequence, two VMs, which are located on different servers, communicate with each other by means of payloads in virtual Ethernet frames, which are transmitted via Berkeley Sockets and TCP/IP in the Guest OS. In the Host OS, we used TAP [24], OpenVSwitch, GRE, IP, IPoIB, TUN, and OFED verbs [27]. This creates significant overhead.

## 5. RESULTS AND CONCLUSIONS

The measurements in our cloud have shown that under the given hardware and software configuration the best speed-up and efficiency could be achieved, if the code was executed by the cores of one CPU or by the CPUs of one server. This is explained by the multiple overheads for virtualized communication that are involved otherwise. Furthermore, the measurements have shown that it is not sufficient to replace the Ethernet network in a standard cloud by Infiniband. Other improvements must be added as well, otherwise a 40 Gbits/s Infiniband can be even slower than a 1 Gbits/s Ethernet. Furthermore, in case of HPC code distribution to different cloud servers it was not possible for the servers to compensate for the resulting communication overhead, because modern hardware accelerator for virtualized communication, such as SR-IOV, were not used by us. Because of that, vCPU data multiplexing and VM data switching was accomplished by OpenVSwitch. The communication overhead made it also impossible to use the remote DMA feature of Infiniband, because multiple extra protocols, device drivers and interfaces had to be added. Our first conclusion is that it is not possible to use clouds with old servers or with standard servers that do not include the latest hardware accelerators for both, virtualized computation and communication. Otherwise OpenStack +KVM +OpenVSwitch is no efficient combination. According to literature and our own findings, there are some cloud-intrinsic problems that make HPC potentially HPC-inefficient. These problems are: 1.) existing cloud schedulers ignore the needs of HPC tasks. One example for that is that heterogeneous cloud hardware and software with different performance capabilities can easily be coupled in a cloud, but with the consequence that the scheduler allocates parallel subtasks of the same HPC job to hardware of different performance, although the subtasks have the same computational intensity. Another example is that communicating subtasks can be scheduled at different points in time, thereby disabling efficient rendezvous-based communication. A third example is that we could not find-out, whether KVM memory protection provides for fast inter-vCPU and for inter-VM communication. However such a communication is indispensable for Open MPI data exchange via shared cache or share memory. 2.) Multi-tenancy and its consequence, the noisy-neighbour problem, effects that the vCPUs of the same VM and that the VMs of the same parallel job are competing with each other for cloud resources. Or second conclusion is therefore that the improvements, which are listed below, should be added to OpenStack to achieve HPC efficiency. Or third conclusion is that it is also not sufficient to install OpenStack on an existing parallel computer, in order to obtain the benefits clouds have. This will result in faster job execution, but efficiency problems were still not solved. A way-out is what most HPC Cloud providers, such as Nimbix, Sabalcore and Nephoscale are doing: they run HPC load on bare metal and with Infiniband. In addition, companies like the UberCloud are using Docker containers [28] and enhancements of it, which are as flexible as VMs, but very lightweight, in order to reduce overhead. Furthermore, customers can use bigger servers in the cloud with more cores and more main memory as they have at home, and thus run their code faster on the one CPU or on one server as at home.

## 6. OUTLOOK

It is our working hypothesis that it is possible to turn every standard cloud into an HPC-capable tool, provided that several or all of the subsequent improvements are made. Of course, not all of them are new, but nevertheless important, which is why we have listed them below.

1. Replacement of Standard Ethernet for inter-server communication by fast Ethernet or Infiniband of at least 10 Gbit/s (better 40 Gbit/s) that is driven by 8-lane PCIe interfaces as minimum.
2. Engagement of the latest hardware accelerators for virtualized communication and computation not only in the CPU, but also on the server motherboard and in the PCIe peripherals. The accelerator that is mandatory as minimum equipment is either SR-IOV or VT-d [29]. However, both do not come for free and have disadvantages as well with respect to system administration and IT security: they do not allow splitting the real interconnection network of the cloud into separate tenant VLANs or VXLANs. A workaround is to use Infiniband partitions. Unfortunately, partitions cannot be configured by OpenStack Neutron.
3. Proper configuration and activation of BIOS-, Hypervisor and Cloud OS options and flags to fully exploit the aforementioned accelerators. In practice, this can be quite difficult.
4. Avoidance of memory and core over committing by too much virtualization. The amount of VM launches and virtual main memory should be limited and carefully monitored. This prevents from excessive paging in guest and host OSes.
5. Avoidance of nested virtualization, unless proper accelerators are available.
6. Employment of Open MPI because of its efficiency, its automatic selection of the fastest communication paths between processing elements and its reluctance in using TCP/IP.
7. Replacement of TCP/IP in the Guest OSes and of IP in the Host OS by stubs such as the Mellanox Messaging Accelerator MXM [14], [30] or by Myrinet Open-MX [31]. This is possible because OpenStack replaces L3 routing by L2 switching, as long as the cloud is in the same hall or rack.
8. Replacement of the existing Guest OS und Host OS Schedulers by an approach that allows for priorities and that is aware of the cloud hardware, its network topology and the problem noisy neighbours, and that differentiates jobs into classes ranging from low performance to high performance computing.
9. Adding to the cloud OS scheduler a gang scheduling in space that allocates communicating vCPUs and VMs to cores in the same CPU chip or to cores in the same server, in order to improve inter-process communication.
10. Adding to the cloud OS scheduler a gang scheduling in time that that schedules communicating vCPUs simultaneously, in order to allow for rendezvous Host OS via blocking Send/Receive.
11. Adding to the cloud OS scheduler a gang scheduling in capability: IO-intensive HPC jobs should be scheduled to servers with fast peripherals. To accomplish that, a job control language (JCL) is needed for clouds, which informs the scheduler about IO-intensiveness of the jobs.
12. Avoidance of waiting for IO resources by an advanced reservation of peripherals, that should be implemented already in the jobs JCL suggested previously.
13. Enhancement of the cloud OS scheduler by a performance delivery model of the cloud and by a resource consumption model of its clients for predictive resource planning. These demand/offer models should allow for interconnect topology-awareness, server

14. Avoidance of the noisy-neighbour problem by partial batch processing in the cloud. The cloud OS scheduler should make an exclusive allocation of vCPUs to jobs, without any time sharing of cores between jobs and users. The time sharing in the cloud should be restricted to another subset of cores, because it disturbs inter-VM and inter-vCPU communication.
15. Incrementing of the problem sizes to at least 100 000 grid points to make the ratio between computation and communication better.
16. Integration of Infiniband management into OpenStack Neutron to avoid alien tools.
17. Provisioning of two different Infiniband device driver, one that is optimized for short L2 frames, the other that is optimized for long L3 IP V6-Jumbo packets: this allows to adapt to the communication requirements of the various VMs.

**Acknowledgement:**

The project was funded by the Scientific Simulation centre Clausthal-Goettingen (SWZ) under contract #11.4.1.